\documentclass[preprint,prd,showpacs]{revtex4} 
 
\begin{document} 
\input{epsf} 
   

\title{Spatially Averaged Quantum Inequalities Do Not  
Exist in Four-Dimensional Spacetime}

\author{L.H. Ford} 
 \email[Email: ]{ford@cosmos.phy.tufts.edu} 
 \affiliation{Institute of Cosmology, Department of Physics and Astronomy\\ 
         Tufts University, Medford, MA 02155}
\author{ Adam D. Helfer} 
 \email[Email: ]{ adam@math.missouri.edu} 
 \affiliation{Department of Mathematics, University of Missouri,
Columbia, Missouri 65211 }  
\author{Thomas A. Roman} 
 \email[Email: ]{roman@ccsu.edu} 
 \altaffiliation[\protect\\ Permanent address: ]{Department of Physics and Earth Sciences, 
         Central Connecticut State University, New Britain, CT 06050 } 
 \affiliation{Institute of Cosmology, Department of Physics and Astronomy\\ 
         Tufts University, Medford, MA~02155} 
 
\begin{abstract} 
We construct a particular class of quantum states for a massless,  
minimally coupled free scalar field 
which are of the form of a superposition of the vacuum and multi-mode 
two-particle states. These states can exhibit local negative energy 
densities. Furthermore, they can produce an arbitrarily large amount  
of negative energy in a given region of space at a fixed time.  
This class of states thus provides an explicit 
counterexample to the existence of a spatially averaged quantum inequality 
in four-dimensional spacetime.  
\end{abstract}


\maketitle 
 
\baselineskip=14pt 
 
\section{Introduction} 
\label{sec:intro} 
 
It is well known that quantum fields can produce  
local renormalized negative energy 
densities. This is despite the fact that the classical expression for  
the energy density appears to be positive definite. The negative energy 
density is possible because renormalization involves an infinite subtraction,  
and is needed to make the stress tensor operator well-defined.  
If there were no restrictions on the  
sorts of negative energy densities attainable, one could  
have a number of bizarre possibilities, including traversable  
wormholes \cite{MT,MTY}, faster-than-light travel \cite{A,K,Olum}  
time travel \cite{MT,MTY,E}, and violations of the second 
law of thermodynamics \cite{F78}. Such phenomena are, at best, rare. 
 
Fortunately, there are known to be  
severe constraints on the magnitude and duration 
of negative energy fluxes or densities, at least in the case of linear fields.  
These are the quantum inequalities, 
which take the typical form \cite{F91,FR95,FR97,FLAN,PF971,PFGQI,FE} 
\begin{equation} 
\int_{-\infty}^\infty \langle :T_{00}({\bf x},t):\rangle \, F(t)\, dt 
\geq - \frac{c}{t_0^d} \,.                  \label{eq:qi} 
\end{equation} 
Here $\langle :T_{00}({\bf x},t):\rangle$ is the renormalized energy  
density in an arbitrary quantum state, $F(t)$ is a normalized 
sampling function with  
a characteristic width $t_0$, $c$ is a dimensionless constant, usually 
somewhat less than unity, and $d$ is the dimension of spacetime. Most 
of the work on quantum inequalities is for the cases of $d=2$ and $d=4$. 
More precisely, Eq.~(\ref{eq:qi}) is a worldline quantum inequality for 
an inertial observer in Minkowski spacetime, where $t$ is the observer's  
proper time. (Recently, worldline inequalities have been established  
in much greater generality in globally hyperbolic spacetimes  
\cite{Fewster}.) The physical interpretation 
of this constraint is that there is an inverse relation between the 
magnitude of the negative energy and its duration. For example, in four 
dimensions, the observer will not see the negative energy last for a time  
longer than about $|\rho_{max}|^{-\frac{1}{4}}$, where $\rho_{max}$ is the 
maximally negative energy density seen by this observer. This type of 
constraint greatly limits the macroscopic effects of negative energy. 
In particular, macroscopic wormholes or ``warp  
drive'' spacetimes are 
severely constrained \cite{FRWH,PFWD,ER}.  
Furthermore, one can use the worldline inequalities to place some constraints  
on the spatial distribution of negative energy \cite{BFR02}. 
All allowed spatial distributions must be such that Eq.~(\ref{eq:qi}) is  
satisfied for {\it every} inertial observer in Minkowski spacetime.   
 
There is a different type of constraint on the energy density, which is that 
the total energy in any quantum state for a quantum field in boundary-free 
Minkowski spacetime must be non-negative. Let the Hamiltonian operator be 
\begin{equation} 
H = \int d^{d-1} x\, :T_{00}({\bf x},t): \,, 
\end{equation} 
where the integral is taken over all space at a fixed time. The expectation  
value of $H$ in an arbitrary quantum state is  non-negative: 
\begin{equation} 
\langle H \rangle \geq 0 \, , 
\end{equation} 
and $\langle H \rangle=0$ only in the vacuum state. Thus any non-vacuum state 
must have net positive energy, even though there can be local regions with 
negative energy density.   
 
This raises the question of whether there are spatial or spacetime quantum  
inequalities analogous to Eq.~(\ref{eq:qi}), but involving sampling over 
space or space and time. For spacetime averages, there  
is an abstract existence proof for a 
very general class of such inequalities in globally hyperbolic 
spacetimes.  These include, for example, inequalities bounding 
violations of the dominant energy condition \cite{Helfer99}.   
However, at present 
quantitative estimates for these are not known. 
 
One can get some spacetime quantum inequalities by simply averaging 
the worldline inequalities over families of worldlines with spatial 
weighting factors, but in at least some 
cases these are known not to be sharp.  For example, 
stronger constraints were found in two-dimensional Minkowksi space 
\cite{FLAN,Roman97}.  We may express these as 
\begin{equation} 
\int dx\, dt\, F_S(x)\, F_T(t)\, \langle :T_{00}({ x},t):\rangle \, 
\geq -B(x_0,t_0) \,.                \label{eq:stqi} 
\end{equation} 
Here $F_S(x)$ is a spatial sampling function with width $x_0$,  
$F_T(t)$ is a temporal sampling function with width $t_0$, and the 
bound $B(x_0,t_0)$ is finite as long as at least one of $x_0$ and $t_0$ 
is nonzero. The sampling functions are assumed to be each normalized 
to unity:
\begin{equation}
\int dx\, F_S(x) \; = \; \int dt\,  F_T(t) \; = \, 1\,. \label{eq:norm}
\end{equation}
 If the sampling functions are both Lorentzian functions, 
then $B \propto (x_0 +t_0)^{-2}$; if they are both Gaussians, then 
$B \propto (x_0^2 +t_0^2)^{-1}$. In either case, one obtains the corresponding 
worldline inequality when $x_0=0$, and a nontrivial bound on the spatially 
sampled energy density when $t_0=0$.  
One consequence of this is that in two dimensions the spacetime averaged results 
are stronger than the temporally averaged ones, since as $t_0\to 0$ (for 
$x_0>0$) one gets a finite bound. 
 
It is natural to ask if there is a generalization of Eq.~(\ref{eq:stqi}) 
to four-dimensional spacetime, especially one which gives a nontrivial 
bound when the sampling is over space only. One of us \cite{Helfer96}  
has given an argument that this is not the case, and that the spatially 
sampled energy density can be unbounded below.  
Actually, the argument was written out there for a more generic 
situation, that of evolution from 
a Cauchy surface in a general curved spacetime. 
While explicit constructions were given in this case, they were by 
pseudodifferential operator techniques, and they were not translated 
into more conventional quantum field-theoretic terms.  Also, the 
details of the more special case of a surface of constant time in 
Minkowski space were not written out.  So no explicit version of the 
argument in conventional special-relativistic quantum field-theoretic 
terms has yet been given. The purpose of the present paper is to provide such 
an example and to draw as many physical insights as possible from it. 
In the following section, we will provide the explicit construction of a class  
of quantum states for the massless, minimally coupled scalar field,  
with negative energy density, as well as give the 
two-point function and the energy density in this class of states.  
We then show that although this state satisfies Eq.~(\ref{eq:qi}), the 
spatially sampled energy density can be arbitrarily negative. Further 
implications of this example are discussed in Sect.~\ref{sec:diss}.  
We work in units where $\hbar=c=1$. (A related construction has subsequently  
been used, in Ref.~\cite{FewR}, to prove that there are no quantum  
inequalities along null geodesics in four-dimensional  
Minkowski spacetime.)

\section{The Energy Density in a Class of Quantum States} 
\label{sec:state} 
 
\subsection{Characterization of the State} 
\label{sec:char} 
 
The particular quantum states which will be used in this paper are of the 
form of a superposition of the vacuum and a multi-mode two-particle state: 
\begin{equation} 
|\psi \rangle = N \left[ |0\rangle + \sum_{{\bf k}_1,{\bf k}_2}  
c_{{\bf k}_1,{\bf k}_2}\, |{{\bf k}_1,{\bf k}_2}\rangle \right] \,, 
                                           \label{eq:state} 
\end{equation} 
where we assume that the $c_{{\bf k}_1,{\bf k}_2}$'s are symmetric.  
Here $|{{\bf k}_1,{\bf k}_2}\rangle$ is a two-particle state containing 
a pair of particles with momenta ${\bf k}_1$ and ${\bf k}_2$, and the 
normalization factor is 
\begin{equation} 
N = \left[1 + \xi \sum_{{\bf k}_1,{\bf k}_2} |c_{{\bf k}_1,{\bf k}_2}|^2  
                                 \right]^{-\frac{1}{2}} \,, 
\label{eq:norm_a} 
\end{equation}
where $\xi = 2$ if ${\bf k}_1 \not= {\bf k}_2$ and 
$\xi = 1$ if ${\bf k}_1 = {\bf k}_2$.  
Note that each of the two-particle states in Eq.~(\ref{eq:state})  
appears twice, since $|{{\bf k}_1,{\bf k}_2}\rangle =  
|{{\bf k}_2,{\bf k}_1}\rangle$; this leads to the factor of $2$ 
in Eq.~(\ref{eq:norm_a}) when ${\bf k}_1 \not= {\bf k}_2$. 
We can quantize the scalar field in a box of volume $V$, so that the momenta 
are discrete. Later, we will let $V \rightarrow \infty$, so that 
\begin{equation} 
\sum_{\bf k} \rightarrow \frac{V}{(2 \pi)^3}\, \int d^3k \,. 
\end{equation} 
In this limit, we replace the coefficients $c_{{\bf k}_1,{\bf k}_2}$ by 
functions of continuous momenta defined by 
\begin{equation} 
b({\bf k}_1,{\bf k}_2) = \frac{V}{(2 \pi)^3}\,c_{{\bf k}_1,{\bf k}_2}\,. 
                                                  \label{eq:infvol} 
\end{equation} 
The normalization factor then becomes 
\begin{equation} 
N = \left[1 + 2 \int d^3k_1\, d^3k_2\, |b({\bf k}_1,{\bf k}_2)|^2  
              \right]^{-\frac{1}{2}} \,.       \label{eq:state_norm} 
\end{equation} 
 
The quantum state is also characterized by a cutoff $\Lambda$, so that 
$|{\bf k}_1|, |{\bf k}_2| < \Lambda$. Later, we will consider the limit 
in which the cutoff is removed, $\Lambda \rightarrow \infty$. We will 
require that the functional form of $b({\bf k}_1,{\bf k}_2)$ be such that 
the integral in Eq.~(\ref{eq:norm}) remain finite in this limit.   
 
The particular choice of $b({\bf k}_1,{\bf k}_2)$ which we adopt is 
the following: 
\begin{equation} 
b({\bf k}_1,{\bf k}_2) = \chi({\bf k}_1+{\bf k}_2) \,  
   (|{\bf k}_1||{\bf k}_2|)^{\nu -\frac{1}{2}}\, ,    \label{eq:b} 
\end{equation} 
where 
\begin{equation}  
\chi({\bf p}) = \left\{ \begin{array}{ll}  
                        \chi_0 \,, & \mbox{if $|{\bf p}| \leq p_0$}  \\ 
                         0\,,     & \mbox{otherwise}         
                         \end{array}  \right.  
\label{eq:chi}    
\end{equation} 
where $\chi_0$ and $p_0$ are arbitrary constants. (The function  
$\chi({\bf p})$ is taken to have a sharp cutoff merely for  
convenience; it could be made smooth without affecting our essential  
argument.) Thus our state is really a  
four-parameter family of states defined by the parameters $\Lambda$, $\nu$, 
$\chi_0$, and $p_0$. 
We will be especially interested in the limit in which the magnitudes of the 
momenta of both particles become large with $p_0$ fixed. In this limit, the pairs 
of particles have almost exactly opposite momenta.  
 
The integral in Eq.~(\ref{eq:norm}) can now be written as 
\begin{equation} 
\chi_0^2 \,\int d^3k\,\int_{|{\bf p}| \leq p_0} d^3p\,  
(|{\bf k}||{\bf k - p}|)^{2\nu -1} \,, 
\end{equation} 
where ${\bf k} = {\bf k}_1$ and ${\bf p} = {\bf k}_1+{\bf k}_2$. 
In the limit that $\Lambda \rightarrow \infty$, the $k$-integration will 
converge at the upper limit provided that 
\begin{equation} 
\nu < - \frac{1}{4} \,. 
\end{equation} 
This is the condition that our quantum state be normalizable in the  
 $\Lambda \rightarrow \infty$ limit.

\subsection{The Two-Point Function} 
\label{sec:twopt} 
 
Now we wish to calculate the form of the two-point function for the massless  
scalar field in the states we have selected. First expand the field  
operator in terms 
of mode functions and creation and annihilation operators as 
\begin{equation} 
\varphi = \sum_{\bf k} (a_{\bf k} f_{\bf k} + a^\dagger_{\bf k} f^*_{\bf k}) 
\end{equation} 
where the mode functions are 
\begin{equation} 
f_{\bf k} = \frac{1}{\sqrt{2 \omega V}}\, 
 {\rm e}^{i({\bf k}\cdot{\bf x} - \omega\, t)} \, ,  \label{eq:modefnt} 
\end{equation} 
with $\omega =|{\bf k}|$. 
 
We are interested in the renormalized two-point function in the state 
$|\psi\rangle$, defined by 
\begin{equation} 
\langle :\varphi(x)\, \varphi(x'):\rangle =  
\langle \psi|\varphi(x)\, \varphi(x') |\psi\rangle - 
\langle 0|\varphi(x)\, \varphi(x') |0\rangle \,. 
\end{equation} 
It may be expressed as 
\begin{eqnarray}  
\langle :\varphi(x)\, \varphi(x'):\rangle &=& \sum_{{\bf k},{\bf k'}} \Bigl[ 
f^*_{\bf k}(x) f_{\bf k'}(x')\, \langle a^\dagger_{\bf k} a_{\bf k'}\rangle 
+f_{\bf k}(x) f^*_{\bf k'}(x')\, \langle a^\dagger_{\bf k'} a_{\bf k}\rangle  
                                  \nonumber \\    
&+& f_{\bf k}(x) f_{\bf k'}(x')\, \langle a_{\bf k} a_{\bf k'}\rangle 
+f^*_{\bf k}(x) f^*_{\bf k'}(x')\,  
\langle a^\dagger_{\bf k} a^\dagger_{\bf k'}\rangle  \Bigr] \, , 
\end{eqnarray}  
where the expectation values are in the state $|\psi\rangle$. 
 
If we evaluate this expression explicitly using the form of the quantum state 
in Eq.~(\ref{eq:state}) and of the mode functions in Eq.~(\ref{eq:modefnt}), and  
then take the infinite volume limit, using Eq.~(\ref{eq:infvol}), the result 
is 
\begin{eqnarray}  
& &\langle :\varphi(x)\, \varphi(x'):\rangle = \frac{2 N^2}{(2 \pi)^3} \, 
{\rm Re} \int  d^3k\, d^3k'\, \frac{1}{\sqrt{\omega \omega'}}\, 
                                                       \nonumber \\ 
&\times&\Bigl[ 2 {\rm e}^{i({\bf k'}\cdot{\bf x'}-{\bf k}\cdot{\bf x})} 
{\rm e}^{i(\omega\, t - \omega'\, t')} \,  
\int d^3k_1 b^*({\bf k}_1,{\bf k}) b({\bf k}_1,{\bf k'}) \nonumber \\ 
 &+&  {\rm e}^{i({\bf k}\cdot{\bf x}+{\bf k'}\cdot{\bf x'})} 
{\rm e}^{-i(\omega\, t + \omega'\, t')}\, b({\bf k},{\bf k'}) \Bigr] \,. 
                                                      \label{eq:2pt} 
\end{eqnarray}  
The first term on the righthand side, which is quadratic in $b$, comes 
from the expectation value in the two-particle component of $|\psi\rangle$ 
alone, whereas the second term,  which is linear in $b$, comes from 
a cross term involving both the vacuum and the two-particle component. 
It is the latter part which will be of greater interest to us. 
So long as $\Lambda < \infty$, the integrals in  Eq.~(\ref{eq:2pt})  
exist and define smooth functions of $x$ and $x'$, so the two-point  
function has the Hadamard form.  
More precisely, the behavior as $x \to x'$ has the Hadamard form.
The falloff as the separation between $x$ and $x'$ increases will be
slow, on account of the sharp cutoff in $\chi (p)$. Had we used a modified 
$\chi$, falling off smoothly, we could have arranged for a more rapid falloff 
as the spatial separation of $x$ and $x'$ becomes large.

\subsection{The Energy Density} 
\label{sec:density} 
 
The energy density in a given quantum state may be obtained from the 
renormalized two-point function. For the massless, minimally coupled  
scalar field, we can write 
\begin{equation} 
\rho = \frac{1}{2}  
\lim_{{{\bf x'}\rightarrow {\bf x}}\atop {t'\rightarrow t}}\left[ 
(\partial_t \partial_{t'} + \mbox{\boldmath $\nabla \cdot \nabla'$})\, 
 \langle :\varphi(x)\, \varphi(x'):\rangle \right] \,. 
\end{equation} 
We may now use Eq.~(\ref{eq:2pt}) to write the energy density as 
\begin{eqnarray} 
\rho &=& \frac{ N^2}{(2 \pi)^3} \, {\rm Re} \int  d^3k\, d^3k'\, 
\sqrt{\omega \omega'} \, (1+ \hat{{\bf k}}\cdot\hat{{\bf k}}')\, 
\Bigl[ 2 {\rm e}^{i({\bf k'}-{\bf k})\cdot{\bf x}} 
{\rm e}^{i(\omega  - \omega')t} \,  
\int d^3k_1 b^*({\bf k}_1,{\bf k}) b({\bf k}_1,{\bf k'})  
                                  \nonumber \\     
&-& {\rm e}^{i({\bf k}+{\bf k'})\cdot{\bf x}} {\rm e}^{-i(\omega + \omega') t} 
\, b({\bf k},{\bf k'}) \Bigr] \,,   \label{eq:rho} 
\end{eqnarray}  
where $\hat{{\bf k}}$ and $\hat{{\bf k}}'$ are unit vectors in the directions 
of ${\bf k}$ and ${\bf k'}$, respectively. 
Let $\rho_1$ denote the term quadratic in $b$, and $\rho_2$ denote the term  
linear in $b$.

Let us first examine the behavior of $\rho_2$.  
If we use the explicit form for $b$ given in Eq.~(\ref{eq:b}), and let 
${\bf p} = {\bf k}+{\bf k'}$, we find 
\begin{equation} 
\rho_2 = -\frac{ \chi_0 N^2}{(2 \pi)^3} \, {\rm Re}\int d^3k\, 
\int_{|{\bf p}|\leq p_0} d^3p\, \sqrt{\omega \omega'} \,  
(1+ \hat{{\bf k}}\cdot\hat{{\bf k}}')\,{\rm e}^{i{\bf p}\cdot{\bf x}}\, 
{\rm e}^{-i(\omega + |{\bf p}-{\bf k}|)t}\, 
  (\omega |{\bf p}-{\bf k}|)^{\nu -\frac{1}{2}} \,. 
\label{eq:rho2_0} 
\end{equation} 
In the limit that $\omega$ becomes large compared to $p_0$, we can expand 
the product $\hat{{\bf k}}\cdot\hat{{\bf k}}'$ as follows: 
\begin{eqnarray} 
\hat{{\bf k}}\cdot\hat{{\bf k}}' &=&  
\frac{\hat{{\bf k}}\cdot({\bf p}-{\bf k})} 
{\sqrt{({\bf p}-{\bf k})\cdot({\bf p}-{\bf k})}} 
= \left( -1 + \frac{\hat{{\bf k}}\cdot{\bf p}}{\omega} \right) 
\left[1 -2\frac{\hat{{\bf k}}\cdot{\bf p}}{\omega} +\frac{p^2}{\omega^2}  
                                                          \right]^{-\frac{1}{2}} 
\nonumber  \\  
&\approx& -1 + \frac{p^2 -(\hat{{\bf k}}\cdot{\bf p})^2}{2 \omega^2} + \cdots \,. 
\end{eqnarray} 
To leading order when $\omega \gg p_0$, we can write $\omega' \approx \omega$. 
If the integral for $\rho_2$ is dominated by values of  $\omega$ large compared  
to $p_0$, we have 
\begin{equation} 
\rho_2 \approx -\frac{\chi_0 N^2}{2 (2 \pi)^3} \, {\rm Re}\int d^3k\, 
\int_{|{\bf p}|\leq p_0} d^3p\,\, \omega^{2(\nu-1)} \, 
[p^2 -(\hat{{\bf k}}\cdot{\bf p})^2] 
\, {\rm e}^{i{\bf p}\cdot{\bf x}}\, {\rm e}^{-2 i \omega t} \,. 
                                                   \label{eq:rho2a} 
\end{equation}

We next perform the angular integrations. If we integrate over the directions 
of ${\bf k}$ with ${\bf p}$ fixed, we find 
\begin{equation} 
\int d^3k\, p^2 = 4 \pi p^2 \int d\omega \, \omega^2 
\end{equation} 
and 
\begin{equation} 
\int d^3k \,(\hat{{\bf k}}\cdot{\bf p})^2 = 
 2 \pi p^2 \int d\omega \, \omega^2\, \int_{-1}^1 dc \, c^2 
= \frac{4 \pi}{3} p^2 \int d\omega \, \omega^2 \, , 
\end{equation} 
where $c = \cos\theta$ and $\theta$ is the angle between  ${\bf k}$ and  
${\bf p}$. Next we integrate over the directions of ${\bf p}$  with  
${\bf x}$ fixed to find 
\begin{eqnarray} 
f_2(p_0,r) &\equiv&  
\int_{|{\bf p}|\leq p_0} d^3p\, p^2\,{\rm e}^{i{\bf p}\cdot{\bf x}} = 
2 \pi \int_0^{p_0}  dp \, p^4 \int_{-1}^1 dc' \,{\rm e}^{i p r c'}= 
\frac{4\pi}{r} \int_0^{p_0}  dp \, p^3\, \sin (p r) \nonumber \\  
&=&    \frac{4\pi}{r^5}\,  
[3(p_0^2 r^2-2)\sin (p_0 r) - p_0 r(p_0^2 r^2-6)\cos (p_0 r)] 
  \, ,                                        \label{eq:f} 
\end{eqnarray} 
where $r = |{\bf x}|$,  
$c' = \cos\alpha$, and $\alpha$ is the angle between  ${\bf x}$ and  
${\bf p}$. The function $f_2(p_0,r)$ has the following behavior: 
\begin{equation} 
f_2(p_0,r) \sim \left\{ \begin{array}{ll}  
                        -(4 \pi /r^2) \,{p_0}^3 \, {\rm cos} \, p_0 r \,,  
                      & \mbox{$r \gg {p_0}^{-1}$}  \\ 
                         (4\pi/5) \, p_0^5 \,, & \mbox{$ r \ll {p_0}^{-1}$}         
                         \end{array}  \right. \,. 
\label{eq:f2sim} 
\end{equation} 
Now we can write Eq.~(\ref{eq:rho2a}) as 
\begin{equation} 
\rho_2 \approx -\frac{\chi_0 N^2}{6 \pi^2} \,f_2(p_0,r)\,  
 \int_{q_2 p_0}^{\Lambda} d\omega \, \omega^{2\nu}\, \cos (2 \omega t) \,, 
                                                   \label{eq:rho2b} 
\end{equation} 
where $q_2$ is a constant chosen so that the approximations made in finding  
the integrand ($\omega \gg p_0$) are valid throughout the range of integration.  
Note that if $t=0$ and $\nu \geq -\frac{1}{2}$, the integral will diverge in 
the $\Lambda \rightarrow \infty$ limit.

Let us now set $t=0$ and $\nu=-1/2$. Then the $\omega$-integral becomes  
\begin{equation} 
\int_{q_2 p_0}^{\Lambda} \frac{d\omega}{\omega} = \ln \left( 
\frac{\Lambda}{q_2 p_0} \right) \,, 
\end{equation} 
so that  
\begin{equation} 
\rho_2 \approx -\frac{\chi_0 N^2}{6 \pi^2} \,f_2(p_0,r)\,  
\ln \left( 
\frac{\Lambda}{q_2 p_0} \right) \,. 
\label{eq:rho2c} 
\end{equation}

We now examine the contribution of $\rho_1$. From  
Eq.~(\ref{eq:rho}), we have  
\begin{equation} 
\rho_1  = \frac{ N^2}{(2 \pi)^3} \, {\rm Re} \int  d^3k\, d^3k'\, 
\sqrt{\omega \omega'} \, (1+ \hat{{\bf k}}\cdot\hat{{\bf k}}')\, 
\Bigl[ 2 {\rm e}^{i({\bf k'}-{\bf k})\cdot{\bf x}} 
{\rm e}^{i(\omega  - \omega')t} \,  
\int d^3k_1 b^*({\bf k}_1,{\bf k}) b({\bf k}_1,{\bf k'})\Bigr] \,. 
\label{eq:rho_1} 
\end{equation} 
If we use Eqs.~(\ref{eq:b}) and ~(\ref{eq:chi}), letting  
${\bf p} = {\bf k}_1+{\bf k}$ and ${\bf p}' = {\bf k}_1+{\bf k}'$,  
and interchanging the order of integration, we can write  
\begin{eqnarray} 
\rho_1  &=& \frac{2\,{\chi_0}^2 \, N^2}{(2 \pi)^3} \,  
{\rm Re} \int d^3k_1\, {|k_1|}^{2 \nu -1} \,  
\int_{|{\bf p}|\leq p_0} d^3p\, \int_{|{\bf p}'|\leq p_0} d^3p'\,  
(1+ \hat{{\bf k}}\cdot\hat{{\bf k}}')\, 
\nonumber \\ 
&\times &{(|{\bf p}-{\bf k}_1|)}^{\nu} \,{|{\bf p}'-{\bf k}_1|)}^{\nu} \, 
{\rm e}^{-i({\bf p}-{\bf p}')\cdot{\bf x}}  
\, {\rm e}^{i[(|{\bf p}-{\bf k}_1|)t-(|{\bf p}'-{\bf k}_1|)t]}\,. 
\label{eq:rho_1_2} 
\end{eqnarray} 
If the ${\bf k}_1$-integral is dominated by values of $|{\bf k}_1| \gg p_0$,  
then we can use ${|{\bf p}-{\bf k}_1|}^{\nu} \, \approx  
{|{\bf p}'-{\bf k}_1|}^{\nu} \approx {|{\bf k}_1|}^{\nu}$. Therefore,  
at $t=0$ we have  
\begin{equation} 
\rho_1  \approx \frac{2 \,{\chi_0}^2 \, N^2}{(2 \pi)^3} \,  
{\rm Re} \int d^3k_1\, {|k_1|}^{4 \nu -1} \,  
\int_{|{\bf p}|\leq p_0} d^3p\, \int_{|{\bf p}'|\leq p_0} d^3p'\,  
 \, (1+ \hat{{\bf k}}\cdot\hat{{\bf k}}') \, 
{\rm e}^{-i({\bf p}-{\bf p}')\cdot{\bf x}} \,. 
\label{eq:rho_1_approx} 
\end{equation} 
Note that in the high $k_1$ limit,  
\begin{equation} 
\hat{{\bf k}}\cdot\hat{{\bf k}}' =  
\frac{({\bf p}-{\bf k}_1) \cdot({\bf p}'-{\bf k}_1)} 
{\sqrt{({\bf p}-{\bf k}_1)\cdot({\bf p}-{\bf k}_1)} \, 
\sqrt{({\bf p}'-{\bf k}_1)\cdot({\bf p}'-{\bf k}_1)}}  
\,\, \rightarrow \, 1 \,. 
\end{equation} 
 
As before, we now set $\nu=-1/2$. A similar calculation  
to that given previously for $\rho_2$ yields  
\begin{equation} 
\rho_1 \approx \frac{2 \, {\chi_0}^2 N^2}{\pi^2} \,f_1(p_0,r)\,  
\ln \left( 
\frac{\Lambda}{q_1 p_0} \right) \,, 
\end{equation} 
where $q_1$ is a constant defined analogously to $q_2$, and  
\begin{equation} 
f_1(p_0,r) \equiv \frac{{(4\pi)}^2}{r^6}\,  
{[\sin (p_0 r) - p_0 r \,\cos (p_0 r)]}^2 \, , 
\label{eq:f1} 
\end{equation} 
and  
\begin{equation} 
f_1(p_0,r) \sim \left\{ \begin{array}{ll}  
                        {(4 \pi p_0)}^2 /r^4 \, {\rm cos}^2 \, (p_0 r) \,,  
                      & \mbox{$r \gg {p_0}^{-1}$}  \\ 
                         {(4\pi /3)}^2 \, p_0^6 \,, & \mbox{$ r \ll {p_0}^{-1}$}         
                         \end{array}  \right. \,. 
\label{eq:f1sim} 
\end{equation} 
Both $|\rho_2|$ and $\rho_1$ diverge logarithmically, but we can arrange for  
$|\rho_2| / \rho_1 > 1$, provided that  
\begin{equation} 
\chi_0 < \frac{1}{12} \, \left(\frac{f_2}{f_1}\right) \, Y \,, 
\end{equation}  
where  
\begin{equation} 
Y \equiv  \, \left[ \frac{\ln \left( 
\Lambda/q_2 p_0 \right)}{\ln \left( 
\Lambda/q_1 p_0 \right)} \right] 
\rightarrow \, 1 \,, \,\,\,\,\,\,\,\,\,\mbox{\rm as \,\,\,\,\,  
$\Lambda \rightarrow \infty$} \,. 
\end{equation} 
This is possible because $\rho_1$ and $\rho_2$ involve different powers  
of $\chi_0$, stemming from the fact that while $\rho_1$ is quadratic in $b$,  
$\rho_2$ is only linear in $b$. 
(We note in passing that the choice of $\nu=-1/2$ gives $\chi_0$  
dimensions of length.) 
 
For any value of $p_0$, the energy density is  
approximately constant in space over a region of size, $r \ll {p_0}^{-1}$.  
In this region, $(f_2/f_1) \approx  
9/(20 \, \pi \, p_0)$. Choose  
\begin{equation} 
\chi_0 < \frac{1}{12} \, \left(\frac{f_2}{f_1}\right) \, Y  
\, \approx \frac{3}{80 \, \pi \, p_0} \,. 
\end{equation}  
where we have used the fact that $Y \rightarrow \, 1$,  
as $\Lambda \rightarrow \infty$. This allows $\rho_2$ to dominate $\rho_1$,  
so that $\rho = \rho_1 + \rho_2$ can be made arbitrarily negative  
over an arbitrarily large region. In summary, we have a normalizable state  
in which the local energy density may be  
made arbitrarily negative at time $t=0$ in a finite region of space.  
It should be pointed out that in the strict  
$\Lambda \rightarrow \infty$ limit, our state will no  
longer have the Hadamard form. However, the key point here is that we  
can consider a sequence of states, each with a finite value  
of $\Lambda$. Therefore each state in this sequence will have the Hadamard form.  
By progressively increasing the values of $\Lambda$ as we vary over  
the states in the sequence, we can make the energy density  
in our spatially sampled region as negative as we like.  
In the next section, we will discuss some of the insights  
which may be drawn from this example.

\section{Implications and Discussion} 
\label{sec:diss}

We are now in a position to draw a number of conclusions from our example. 
The first conclusion is the proof that there are no lower bounds on the spatially  
sampled energy density for the massless minimally coupled scalar field  
in four spacetime dimensions. To be more precise, 
consider the spatial average of the energy density 
\begin{equation} 
S(F) = \int d^3x\, F({\bf x})\, \langle :T_{00}({\bf x},t):\rangle 
\end{equation} 
at time $t$, where $F({\bf x})$ is a spatial sampling function. Let $r_0$ 
be the characteristic width of this function. More generally, $r_0$ can 
be the maximum width of $F({\bf x})$ in any direction. If there were to be 
spatial quantum inequalities here, $S(F)$ would have to have a lower bound 
when we vary over all quantum states for a given $F({\bf x})$. However, the 
energy density in the above example varies in space only over a scale 
of the order of 
\begin{equation} 
\lambda_0 = \frac{2 \pi}{p_0}\, , 
\end{equation} 
which can be arbitrary. Furthermore, we can make the energy density 
in a region whose size is less than $\lambda_0$ arbitrarily negative.  
For a given sampling function $F({\bf x})$, we can choose $\lambda_0 > r_0$, 
so that the energy density is approximately constant over the region being  
sampled and the spatial sampling has no effect, yet the energy density in that  
region is arbitrarily negative. Thus $S(F)$ is unbounded below. However, this 
state satisfies the temporal quantum inequality, Eq.~(\ref{eq:qi}). One  
can argue from energy conservation that physically what is happening is that  
there must be large fluxes of positive energy entering the region, which  
damp out the negative energy sufficiently for the temporal inequality to hold.  
Recall from Eqs.~(\ref{eq:rho2_0}) and ~(\ref{eq:rho_1_2}) that $\rho_2$ oscillates  
much more rapidly in time than does $\rho_1$. Thus if $\rho_2 < 0$ at one time,  
it has changed sign a short time later. This causes the time integral of $\rho_2$  
to tend to average to zero, and allows the worldline quantum inequalities to  
be satisfied.

\begin{figure} 
\begin{center} 
\leavevmode\epsfysize=3cm\epsffile{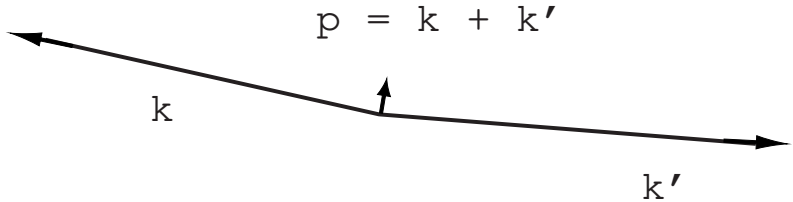} 
\end{center} 
\caption{The pairs of particles in 
the two-particle component of the state have nearly but not exactly opposite 
momenta, in the high frequency limit. Although their momenta nearly cancel,  
their energies do not.} 
\label{fig:oppks} 
\end{figure}

We can gain more physical insight into the behavior of this class of states 
by recalling that in the high frequency limit, the pairs of particles in 
the two-particle component of the state have nearly but not exactly opposite 
momenta, as shown in Fig.~\ref{fig:oppks}.  This allows the scale for the  
spatial variation of $\rho$ to be large compared to the temporal scale, which  
is set by the characteristic frequency of the excited modes. This can be seen in 
Eq.~(\ref{eq:rho2a}), where the spatial factor ${\rm e}^{i{\bf p}\cdot{\bf x}}$ 
oscillates much more slowly than the temporal factor ${\rm e}^{-2 i \omega t}$. 
One can summarize the difference between the spatial and temporal behaviors 
by noting that it is possible for the three-momenta of a pair of particles to  
cancel, but not their energies. We can also see that it is a very special 
property of two-dimensional spacetime which allows for the existence of 
spatial bounds. Here the momenta of a pair of particles must be either exactly  
parallel or exactly anti-parallel. However, if they are anti-parallel, 
there is a factor analogous to the factor of  
$1+ \hat{{\bf k}}\cdot\hat{{\bf k}}'$ in Eq.~(\ref{eq:rho}) which suppresses  
their contribution. Although in this paper we have looked only at  
four-dimensional spacetime, it seems likely that one can construct similar 
counterexamples to spatial quantum inequalities in all spacetime dimensions 
greater than two.

We can also understand why the Hamiltonian, obtained by integrating over  
all space, is bounded below, yet the spatial average over any sampling 
function with finite width is not. The integration over all space removes  
the contribution of the $\rho_2$ term, leaving only the non-negative 
contribution of the $\rho_1$ term. Thus the net energy in any state must 
be non-negative. However, it is possible for the spatial distribution at 
a fixed time to be such that the compensating positive energy is arbitrarily far 
from the negative energy, and hence violates a spatial quantum inequality. 
 
The current results also shed light on an earlier proposed explanation  
of the ``Garfinkle box'', given in Ref.~\cite{BFR02}.  
This refers to an unpublished result of Garfinkle \cite{GAR},  
who showed that the total energy of a scalar field, $E$, contained  
within an imaginary box in Minkowski spacetime, at fixed time, is unbounded below.  
The box is ``imaginary'' in the sense that there are no physical boundaries  
at the walls of the box. The Garfinkle result is that there  
exist quantum states for which $E$ is  
arbitrarily negative. The proposed explanation given in Ref.~\cite{BFR02}  
for the unboundedness of the total energy $E$ in this box involved two factors:  
(1) The energy is measured at a precise instant in time, and (2) the walls of  
the box are sharply defined. It was argued that this allows an arbitrary  
amount of negative energy to have entered the box by time $t$, while at  
this time excluding an even larger amount of positive energy which may be  
just outside of the box at time $t$. The results of the present paper  
show that, although this is {\it an} explanation of the Garfinkle result, it  
is not the only explanation. In particular, the unboundedness of the  
energy in our case does not require the spatial region over which we  
average to be sharply defined. 
 
In this connection, we emphasize that our results show 
that the unboundedly negative energies cannot be ascribed to ``edge effects'' 
associated with the sampling function.  This is because in our examples, the 
energy density is uniformly negative throughout the sampling region. 
 
While spatially averaged quantum inequalities do not exist, we noted earlier 
that 
spacetime averaged ones do. This means that the lower bounds associated with the 
spacetime averages must diverge as the temporal averaging scale goes to zero.  
It is natural to conjecture that this is related to the divergence in the 
spatially averaged inequalities, as follows.  Suppose the most severe divergence 
attainable in the spatial average is 
\begin{equation} 
\int dx\, F_S({\bf x})\, \langle :T_{00}({\bf x},t):\rangle  
  \qquad\hbox{diverges as}\quad -f(\Lambda )\quad\hbox{as} 
  \quad \Lambda\to \infty \,, 
\end{equation} 
for some function $f$ of the ultraviolet cutoff $\Lambda$.  Similarly, suppose 
that for a temporal averaging function $F_T(t)$ with 
characteristic width $t_0$, we have 
\begin{equation} 
\int dx\, dt\, F_S({\bf x})\, F_T(t)\, \langle :T_{00}({\bf x},t):\rangle  
  \qquad\hbox{diverges as}\quad -g(t_0)\quad\hbox{as}\quad t_0\to 0  \,,           
\end{equation} 
for some function $g(t_0)$.  It is natural to ask if $f(\Lambda )\sim 
g(1/\Lambda )$.  A relation like this would probably give the most direct 
physical significance to the divergences in the spatial averages.  (Such a 
relation would not hold in two dimensions, but might in four.) 
 
The logarithmic divergences (of the energy density with the cutoff $\Lambda$) 
are rather slow.  We do not know whether it is possible to construct examples in 
Minkowski space with more severe divergences.  (In Ref.~\cite{Helfer96}, more 
severe divergences were found at Cauchy surfaces with non-zero second 
fundamental forms.  In generic curved spacetimes, this would be the case for all 
Cauchy surfaces.) 
 
In summary, we have shown that there are no purely spatially averaged  
quantum inequalities over bounded regions in four-dimensional 
Minkowski spacetime, even though the integral over all space is bounded.
  However,  
spacetime averaged quantum inequalities do exist. As noted earlier,  
in at least the two-dimensional case, 
spacetime averaging seems to lead to tighter bounds than does 
temporal averaging alone. Recent numerical results of 
Dawson and Fewster \cite{DF} indicate that similar behavior occurs in  
four dimensions. However, at present the extent to which spacetime  
averaging improves temporal averaging is unclear.  
Thus there appear to be two approaches for better  
understanding the allowed spatial distributions of negative energy. One is  
the search for spacetime averaged quantum inequalities, and the other is the  
continuation of the program begun in Ref.~\cite{BFR02}, which seeks to  
extract as much information as possible from the worldline inequalities.  
Both of these approaches are currently under investigation. 
 
\vskip 0.2in    
 
\centerline{\bf Acknowledgments}   
 
The authors would like to thank Chris Fewster for extensive discussions  
on these issues, and Bill Unruh for useful comments.  
TAR is grateful to the Tufts Institute of Cosmology, to Werner Israel and  
the Physics and Astronomy Department of the University of Victoria, and to  
the Mathematical Physics group at the University of York for their  
hospitality during this work. LHF and TAR would also like to  
thank the Erwin Schr{\"o}dinger Institute in Vienna for  
hospitality while this work was completed. This research was supported  
in part by NSF Grants No. Phy-9800965 (to LHF) and  No. Phy-9988464 (to TAR).

\end{document}